\documentclass[useAMS,  usenatbib]{mn2e}
\usepackage{amsmath}
\usepackage{cases}
\usepackage{txfonts}
\usepackage{tipa}
\usepackage{amsfonts}
\usepackage{amssymb}
\usepackage{graphicx}

\title[The stellar metallicity distribution of the Milky Way from the BATC survey]{The stellar metallicity distribution of the Milky Way from the BATC survey}

\author[Xiyan Peng et al.]{Xiyan Peng$^{1,2,3}$,
 Cuihua Du$^{3}$ \thanks{E-mail:ducuihua@ucas.ac.cn(Du); zywu@nao.cas.cn(Wu); xypeng@bao.ac.cn(Peng)},
Zhenyu Wu$^{1,2\star}$, Jun Ma$^{1,2}$, Xu Zhou$^{1,2}$ \\
$^{1}$National Astronomical Observatories, Chinese
Academy of Sciences, Beijing 100012, P. R. China\\
$^{2}$Key Laboratory of Optical Astronomy, National Astronomical Observatories, Chinese Academy of Sciences, Beijing, 100012, China\\
$^{3}$School of physics, University of the Chinese Academy of
Sciences, Beijing 100049, P. R. China}

\begin{document}

\date{Received}

\pagerange{\pageref{firstpage}--\pageref{lastpage}} \pubyear{2002}

\maketitle

\label{firstpage}

\begin{abstract}
Using the stellar atmospheric parameters such as effective
temperature and metallicity derived from SDSS spectra for $\sim2200$
main sequence (MS) stars which were also observed by Beijing -
Arizona - Taiwan - Connecticut (BATC) photometric system, we develop
the polynomial photometric calibration method to evaluate the
stellar effective temperature and metallicity for BATC multi-color
photometric data. This calibration method has been applied to about
160 000 MS stars from 67 BATC observed fields. Those stars have
colors and magnitudes in the ranges $0.1<d-h<1.4$ and $14.0\leq
d<21.0$. We find that there is a peak of metallicity distribution at
$\rm{[Fe/H]}\sim-1.5$ in the distance from the Galactic plane $|Z|>
5~\rm{kpc}$ which corresponds to the halo component and a peak at
$\rm{[Fe/H]} \sim -0.7$ in the region 2 $<|Z|\leqslant  5~\rm{kpc}$
where is dominated by the thick disk stars. The mean stellar
metallicity smoothly decreases from $-0.65$ to $-0.78$ in the
interval $0.5 < |Z| \leqslant  2~ \rm{kpc}$. Metallicity
distributions in the halo and the thick disk seem invariant with the
distance from the Galactic plane.
\end{abstract}


\section{INTRODUCTION}
Studies of the Galactic stellar chemical distribution are imperative
for the Galaxy formation and evolution. The metal-poor end of the
metallicity distribution function reveals information about the
early era of star formation in the Galaxy, while the metal-rich end
of the metallicity distribution function reflects recent Galaxy
conditions \citep{sch55}. There is a wide range of Galaxy formation
models corresponding to different predications about the stellar
metallicity distribution. \citet{egg62} provided a steady Galaxy
formation scenario, ELS model. At the beginning of the Galaxy
formation, a relatively rapid radical collapse of the initial
protogalactic cloud happened, followed by the equally rapid setting
of gas into rotating disk. A steady process of the Galaxy formation
leads to smooth distribution of stars. However, recent discoveries
of complex substructure conflict with the simple collapse model.
Many irregular substructures such as Sagittarius dwarf tidal stream
\citep{ive00,yan00,viv01,maj03}, Virgo Stellar Stream
\citep{duf06,pri09} and the Monoceros stream closer to the Galactic
plane \citep{new02,roc03} were found in the recent large area sky
surveys. Recent observations show that the Galactic structure is a
complex and dynamic structure. The SZ model that the Galaxy formed
at least partly through the merge of larger systems was advanced by
\citet{sea78}. This hypothesis is supported by recent studies of
stellar kinematics and metallicity distribution
\citep{maj96,bel06,jur08}.

The metallicity distribution of different Galactic components can be
used to constrain the Galaxy formation and evolution models. The
thin disk, the thick disk and the halo are the basic components of
the Galaxy. Stars on more radial orbits are metal-poor than stars on
planar orbits. This conclusion is supported by the observation data
\citep{bah80,gil83,gil89}. The ELS model cannot explain the Galaxy
halo without metallicity gradient while the SZ model is able to
account for the lack of metallicity gradient of the halo. Different
thick disk formation scenarios correspond to different abundance
distribution in the thick disk. If the thick disk was formed in the
Galaxy collapse, there must be a stellar metallicity gradient in the
thick disk. \citet{chen11} found that a vertical gradient of $-0.22
\pm 0.07$ dex $\rm{kpc^{-1}}$ for the thick disk. On the other hand,
the absence of the stellar metallicity gradient in the thick disk
favor the thick disk was formed through merger of larger system or
via kinematic heating of the thin disk. There has been much debate
about the thick disk, some researchers pointed out that the two
components of disk cannot be separable from one another
\citep{nor91,bov12}. \citet{ive08} suggested that the metallicity
and velocity distribution of the disk can be described by a single
disk component instead of two separate ones.

The most precise methods to derive the stellar metallicity are based
on the spectroscopic survey \citep{all06,ive12}. However, imaging
survey data provide better sky and depth coverage than spectroscopic
survey data. The photometric metallicity methods are based on the
theory that the density of metallicity absorption lines is highest
in the blue spectral region. Low metal abundance produces weak
atmospheric absorption lines and increased ultraviolet emission
\citep{sch55,san69,bee05}. \citet{sie09} derived approximate stellar
abundance through measurement of ultraviolet excess, despite this
method is less precise in larger distance. \citet{ive08}
demonstrated that for the MS stars, the stellar metallicity estimate
can be derived through SDSS color indices. This calibration of the
metallicity has been applied for millions of stars to explore the
metallicity distribution of the Galaxy. In this paper, we present an
analysis of stars in the BATC photometric survey. Some Galactic
structure and stellar metallicity distribution studies have been
carried out using the BATC photometric data
\citep{du03,du04,du06,pen12}. \citet{pen12} adopted the spectral
energy distribution (SED) fitting method to evaluate the stellar
metallicity in the Galaxy. With the new metallicity calibration
method and more BATC photometric survey fields used in this paper,
it is possible to derive more information about the stellar
metallicity distribution of the Galaxy.

In this paper, we attempt to study the Galactic stellar metallicity
distribution using photometric metallicity method. In total, there
are 67 BATC photometric survey fields used in this paper. Section 2
of this paper describes the way to select star sample. The methods
of stellar atmospheric parameters estimation are described in
Section 3. The stellar metallicity distribution is discussed in
Section 4. Finally, in Section 5, we summarize our main conclusions in
this study.

\section{OBSERVATIONS AND DATA}
\subsection{BATC survey and data reduction}
The BATC photometric survey is carried out by the BATC multi-color
system with the 60/90 cm f/3 Schmidt telescope at the Xinglong
station, National Astronomical Observatories of Chinese Academy of
Sciences (NAOC). A Ford 4096 $\times$ 4096 CCD camera was equipped
at the prime focus of the telescope. The field of view is 92 $\times
$ 92 arcmin$^2$ with a spatial scale of 1.36 arcsec pixel$^{-1}$.
The BATC system contains 15 intermediate-band filters, covering a
wavelength from 3000 to 10 000 {\AA} which are designed to avoid
strong night-sky emission lines \citep{fan96}. The parameters of the
BATC filters are listed in Table 1. Col. (1) and Col. (2) represent
the ID of the BATC filters, Col. (3) and Col. (4) represent the
central wavelengths and FWHM of the 15 filters, respectively
\citep{pen12}. Using automatic data-processing software, PIPELINE I
\citep{fan96}, we carry out the standard procedures of bias
subtraction and flat-field correction. The PIPELINE II reduction
procedure is performed on each single CCD frame to get the point
spread function (PSF) magnitude of each point source \citep{zho03}.
The BATC magnitude adopt the monochromatic AB magnitude as defined
by \citet{oke83}. The detailed description of the BATC photometric
system and flux calibration of the standard stars can be found in
\citet{fan96} and \citet{zho03}.

% Table 1_______________________________
\begin{table}
\begin{minipage}{120mm}
\caption{Parameters of the BATC filters }
\begin{tabular}{cccc}\hline
\hline
No. & Filter &  Wavelength  & FWHM \\
    &         & (\AA)   & (\AA)   \\
\hline
1  & $a$ & 3371.5  & 359    \\
2  & $b$ & 3906.9  & 291    \\
3  & $c$ & 4193.5  & 309    \\
4  & $d$ & 4540.0  & 332    \\
5  & $e$ & 4925.0  & 374    \\
6  & $f$ & 5266.8  & 344    \\
7  & $g$ & 5789.9  & 289    \\
8  & $h$ & 6073.9  & 308   \\
9  & $i$ & 6655.9  & 491   \\
10 & $j$ & 7057.4  & 238   \\
11 & $k$ & 7546.3  & 192   \\
12 & $m$ & 8023.2  & 255   \\
13 & $n$ & 8484.3  & 167   \\
14 & $o$ & 9182.2  & 247   \\
15 & $p$ & 9738.5  & 275   \\
\hline \hline
\end{tabular}
\end{minipage}
\end{table}

Accurate determination of the properties of the components of the
Galaxy requires surveys with sufficient sky coverage to assess the
overall geometry \citep{de10}. Most previous investigations about
the Galactic metallicity distribution used only one or a few fields
along the lines-of-sight \citep{du04,sie09,kar09}. The fields used
in this paper are toward different directions, the Galactic center,
the anticentre at median and high latitudes,
$20^{\circ}<|b|<60^{\circ}$. We try to choose the MS stars to carry
out our study. The distribution of all stars in the BATC d filter in
the BATC T192 and T193 fields is shown in Figure 1. Owing to the
BATC observing strategy, stars brighter than $d=14$ mag are
saturated. In addition, from Figure 1, we can see that  star counts
are not completed for magnitudes fainter than $d=21.0$ mag. So our
work is restricted to the magnitude range $14\leqslant d < 21$. The
MS stars with colors located in the proper range for application the
photometric metallicity method are needed. The specifical criteria
used to choose MS stars are the following:

1. $14 \leq d < 21$;

2. $0.1 < (d-h) < 1.4$;

3. $0 < (d-n) <  2.5$;

4. $-0.7 < (d-n)-1.6(d-h) < 0.8$.

The $d-h$ versus $d-n$ color-color diagram for all the point sources
with $14 \leq d < 21$ in the BATC T192 and T193 fields is shown in
Figure 2(a). As can be seen from the panel (a), the point sources
show a MS stellar locus but have significant contamination from
giants, manifested in the broader distribution overlying the narrow
stellar locus. \citet{jur08} applied a procedure which rejected
objects at distance larger than 0.3 mag from the stellar locus in
order to remove hot dwarfs, low-red shift quasars, white /red dwarf
and unresolved binaries from their sample. The procedure does well
for the data in the high latitude fields of the SDSS survey
\citep{kar03a,yaz10}. The analogous method is used in our work, the
criteria 2, 3 restrict the color of stars in the proper range for
applying the photometric metallicity method and criterion 4 removes
objects far away from the dominant stellar locus. Figure 2(b) gives
the color-color diagram of the MS stars after rejecting those
objects which lied significantly off the stellar locus. The criteria
1--4 have been applied to BATC photometric data to choose the MS
stars. In total, there are about 160 000 MS stars in the 67 fields
in our study. The Galactic longitude and latitude of 67 BATC
observation fields are shown in Figure 3. Table 2 lists the
locations of the observed fields and their general characteristics.
Column 1 represents the BATC field name, columns 2--4 represent the
right ascension, declination and epoch, columns 5 and 6 represent
the Galactic longitude and latitude, columns 7 and 8 represent the
$E(B-V)$ color excess and the number of MS star of each field used
in this study, respectively. The values of $E(B-V)$ derived from
\citet{sch98} are small for most BATC survey fields.

\begin{figure}
\begin{center}
\includegraphics[width=8cm]{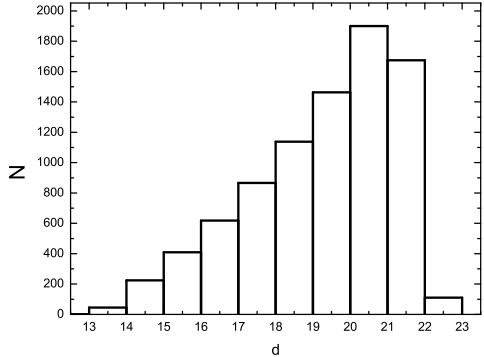}
\caption{The distribution of all stars with the BATC d filter in the
BATC T192 and T193 fields } \label{fig1}
\end{center}
\end{figure}

\subsection{SDSS stellar spectroscopic atmospheric parameters}
The stellar spectroscopic atmospheric parameters are needed to
derive the stellar metallicity and effective temperature estimates
for BATC photometric data. We use the stellar spectroscopic data of
SDSS. The SDSS has been in routine survey since 2000 \citep{yor00}.
This survey is carried out by two instruments: a drift-scan imaging
camera \citep{gun98} and a pair of double spectrographs, fed by 640
optical fibers. Each fiber covers wavelength coverage 3800--9200
{\AA}, spectral resolution of R is about 1800. Signal to noise ratio
is about 10 per pixel at $r_{psf} = 19.5$ \citep{yor00,Sme12}. The
estimates of the precision of the derived SDSS stellar spectroscopic
atmospheric parameters $T_{eff}$, log $g$, [Fe/H] and [$\alpha $
/Fe] are 150 K, 0.23 dex, 0.21 dex and 0.1 dex
\citep{lee08a,lee08b,all08,eis11}. Galactic stellar spectroscopic
data of all common spectral types are available in the SDSS DR8
sppParams table. There are 2769 BATC stars which have been observed
by SDSS spectral observation. The criteria 1--4 listed in the
section 2.1 are also applied to the 2769 stars to choose the MS
stars. Meanwhile, about 6.7\% stars in 2769 stars with log $g< 3$
are excluded to remove giant stars. Finally, about 2200 MS star are
chosen to calibrate photometric metallicity and effective
temperature estimates. The 2200 MS stars cover mainly a range of
spectroscopic metallicity [Fe/H] from $-2$ to 0 and effective
temperature $T_{eff}$ from 4000 K to 7000 K.

% Table 1_______________________________

   \begin{table*}
    \centering

     \caption{Relative information for the BATC
    observation fields}
         \begin{tabular}{ccccccccc}%{p{0.8cm}p{1.4cm}p{1.4cm}p{0.5cm}ccp{0.5cm}p{0.5cm}}
            \hline

             Field & $\rm{R.A.}$  & $\rm{Decl.}$ & $\rm{epoch}$   & $l$(deg)  & $b$(deg)   & $E(B-V)$   & \rm{star number}   \\
  \hline

T003&    00:21:26.90&   16:12:32.0&      1950&  113.48&   -45.89& 0.06& 1291\\
T032&    02:43:55.80&  -00:34:49.5&      1950&  173.10&   -52.13& 0.04& 1348\\
T033&    03:11:38.10&  -03:00:29.0&      1950&  183.69&   -48.12& 0.06& 1437\\
T034&    04:28:11.20&   00:45:25.0&      1950&  194.27&   -30.38& 0.10& 2286\\
T035&    04:34:06.50&  -03:17:23.0&      1950&  199.20&   -31.23& 0.07& 4806\\
T192&    17:49:58.80&   70:09:26.0&      1950&  100.57&    30.64& 0.03& 2986\\
T193&    21:55:34.00&   00:46:13.0&      1950&   59.78&   -39.74& 0.07& 2843\\
T195&    23:02:26.80&   12:03:06.0&      1950&   86.27&   -42.84& 0.15& 1909 \\
T196&    23:06:34.90&   17:54:20.0&      1950&   91.38&   -38.34& 0.12& 1798 \\
T246&    03:02:16.30&  -00:19:47.0&      1950&  178.36&   -48.12& 0.11& 1386 \\
T247&    03:02:16.32&   17:05:22.3&      1950&  162.88&   -35.02& 0.15& 1504 \\
T288&    08:42:30.50&   34:31:54.0&      1950&  188.98&    37.48& 0.03& 1584 \\
T289&    08:46:20.50&   15:40:41.0&      1950&  211.44&    32.90& 0.03& 1992  \\
T290&    09:03:49.95&   17:34:27.9&      1950&  211.22&    37.49& 0.03& 1913 \\
T291&    09:32:00.30&   50:06:42.0&      1950&  167.84&    46.45& 0.01& 1293  \\
T324&    01:31:14.30&   01:20:54.0&      1950&  144.17&   -59.51& 0.03& 1336 \\
T325&    03:07:15.40&   02:22:00.0&      1950&  176.77&   -45.36& 0.09& 1407  \\
T326&    03:45:27.00&   01:30:08.0&      1950&  186.06&   -38.70& 0.15& 1172 \\
T327&    07:51:40.70&   56:23:06.0&      1950&  161.48&    31.06& 0.05& 2155  \\
T328&    09:10:57.30&   56:25:49.0&      1950&  160.25&    41.95& 0.02& 1391 \\
T329&    09:53:13.30&   47:49:00.0&      1950&  169.95&    50.38& 0.01& 1253 \\
T338&    14:42:50.48&   10:11:12.2&      1950&    5.79&    58.17& 0.02& 1704 \\
T343&    23:31:58.50&   02:16:47.0&      1950&   87.93&   -55.00& 0.05& 1411 \\
T349&    09:13:34.60&   07:15:00.5&      1950&  224.10&    35.28& 0.05& 1892 \\
T359&    22:33:51.10&   13:10:46.0&      1950&   79.71&   -37.84& 0.06& 2297    \\
T360&    02:56:31.80&  -00:00:29.0&      1950&   76.51&   -48.93& 0.07& 1346 \\
T362&    10:47:55.40&   04:46:49.0&      1950&  245.67&    53.37& 0.03& 1395   \\
T363&    09:52:00.00&  -01:00:00.0&      1950&  239.32&    38.89& 0.05& 1762 \\
T381&    02:39:38.60&   00:00:06.5&      1950&  171.71&   -51.84& 0.03& 1327   \\
T387&    04:20:43.80&  -01:27:33.0&      1950&   95.29&   -33.14& 0.14& 5232\\
T475&    17:14:53.00&   50:12:36.0&      1950&   76.90&    35.48& 0.02& 2978 \\
T477&    08:45:48.00&   45:01:17.0&      1950&  175.74&    39.17& 0.03& 3038 \\
T491&    22:14:36.10&  -00:02:07.6&      1950&   62.10&   -43.62& 0.07& 2101\\
T506&    17:43:18.30&   30:34:42.9&      1950&   55.36&    26.76& 0.06& 5510 \\
T510&    08:48:24.00&   12:00:00.0&      1950&  215.66&    31.86& 0.02& 2417 \\
T512&    00:41:48.50&   85:03:27.6&      1950&  122.83&    22.46& 0.09& 5179  \\
T515&    17:15:36.00&   43:12:00.0&      1950&   68.35&    34.86& 0.07& 6996 \\
T517&    03:51:43.04&   00:10:01.6&      1950&  188.63&   -38.24& 0.20& 1051 \\
T518&    09:54:05.60&  -00:13:24.4&      1950&  238.93&    39.77& 0.04& 1658    \\
T520&    15:36:17.04&  -00:10:03.4&      1950&     5.6&    41.32& 0.10& 3129\\
T521&    21:39:19.48&   00:11:54.5&      1950&   56.13&   -36.81& 0.06& 3467   \\
T523&    00:40:00.20&   40:59:43.0&      1950&   121.17&  -21.57& 0.14& 5624   \\
T524&    01:31:01.70&   30:24:15.0&      1950&   133.61&  -31.33& 0.06& 4249\\
T525&    01:34:00.70&   15:31:55.0&      1950&   138.62&  -45.70& 0.05& 1193   \\
T527&    07:36:54.52&   65:35:58.3&      1950&   150.72&   29.68& 0.04& 2522\\
T528&    09:55:34.33&   69:21:46.5&      1950&   141.75&   41.16& 0.08& 1598   \\
T533&    14:01:26.70&   54:35:25.0&      1950&   102.04&   59.77& 0.01& 1219\\
T534&    15:14:34.80&   56:30:33.0&      1950&    91.58&   51.09& 0.01& 1505   \\
T601&    22:14:36.10&  -00:02:07.6&      2000&   136.86&   58.66& 0.03& 1198 \\
T702&    22:37:00.64&   34:25:37.1&      2000&    93.72&   20.71& 0.09& 5465 \\
T703&    02:13:58.16&   27:53:06.3&      2000&   144.36&  -31.53& 0.07& 2239  \\
T704&    02:27:15.14&   33:35:12.0&      2000&   144.88&  -25.17& 0.08& 3153 \\
Ta02&    02:56:12.00&   13:23:00.0&      2000&   163.65&  -39.43& 0.17& 3283  \\
Ta04&    17:12:12.00&   64:09:00.0&      2000&    94.05&   34.97& 0.03& 2873 \\
Ta05&    22:23:42.00&   17:06:00.0&      2000&    79.66&  -33.08& 0.05& 3267    \\
Ta06&    23:35:48.00&   26:45:00.0&      2000&   102.71&  -33.13& 0.06& 3024 \\
Ta07&    00:46:26.60&   20:29:23.0&      2000&   121.35&  -42.37& 0.04& 2319    \\
Ta09&    02:57:56.40&   13:00:59.0&      2000&   164.37&  -39.47& 0.17& 1871 \\
Ta11&    04:13:20.70&   10:28:35.0&      2000&   182.41&  -28.30& 0.52& 2074    \\
Ta12&    09:07:17.60&   16:39:53.0&      2000&   212.13&   37.38& 0.03& 2021 \\
Ta17&    15:09:33.20&   07:31:38.0&      2000&     8.45&   51.87& 0.03& 1485    \\
Ta24&    23:24:00.50&   16:49:29.0&      2000&    94.66&  -41.20& 0.03& 2316 \\
Ta28&    08:28:28.51&   30:25:10.5&      2000&   192.74&   33.11& 0.05& 2744   \\
Ta29&    10:58:04.30&   01:29:56.0&      2000&   251.46&   52.65& 0.03& 1200\\
Ta30&    10:32:08.78&   40:12:44.7&      2000&   179.42&   58.46& 0.01& 1907   \\
u121&    03:58:20.59&   00:24:13.4&      2000&   189.28&  -37.36& 0.31& 948  \\
Ta26&    09:19:57.12&   33:44:31.3&      2000&   191.11&   44.42& 0.02& 1724    \\

    \hline
   \end{tabular}

\end{table*}

\section{STELLAR METALLICITY AND EFFECTIVE TEMPERATURE ESTIMATION}

\subsection{Stellar effective temperature estimation}
The stellar effective temperature can be determined from color
index, for example, \citet{ive08} derived the effective temperature
using SDSS $g-r$ color only. \citet{cas06} derived an empirical
effective temperature for G and K dwarfs by applying their own
implementation of the infrared flux method to multiband photometric
data. The spectroscopical effective temperature versus $d-n$ color
diagram of 2200 BATC MS stars is shown in the upper panel of Figure
4. Color $d-n$ is sensitive to the effective temperature. In order
to get more accurate effective temperature, the best fit expression
is divided into two parts. Equation 1a corresponds to $d-n$ in the
range from 0 to 1.5 and equation 1b corresponds to $d-n$ range from
1.5 to 2.5.

\addtocounter{equation}{1}
\begin{align}
\log(T_{eff}/{\rm K})&=3.853-0.132(d-n) \tag{\theequation a} \\
\log(T_{eff}/{\rm K})&=3.715-0.045(d-n) \tag{\theequation b}
\end{align}

Equation (1) has been applied to the 2200 stars to reproduce the
effective temperature. The residuals between the spectroscopic
effective temperature and photometric estimates are shown in the
bottom panel of Figure 4. Compared with spectroscopic effective
temperature, stars with $0.1 < d-h < 1.5$ have residuals of 0.013
dex (corresponding to 170 K) and stars with $1.5 < d-h < 2.5$ have
less residuals 0.005 dex (corresponding to 50 K).

\begin{figure}
\begin{center}
\includegraphics[width=8cm]{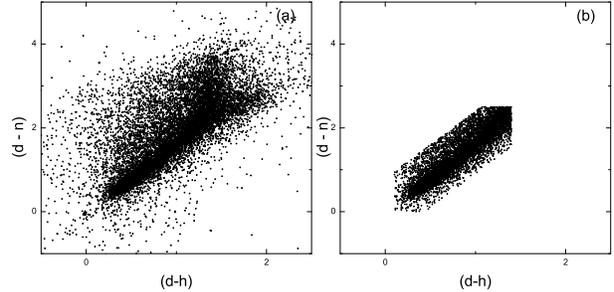}
\caption{The $d-h$ vs. $d-n$ color-color diagrams for the BATC T192
and T193 fields. The panel (a) is the distribution for all point
sources and panel (b) is the color-color diagram for the MS stars in
the proper color range for applying the photometric metallicity
method.} \label{fig1}
\end{center}
\end{figure}

\begin{figure}
\begin{center}
\includegraphics[width=8cm]{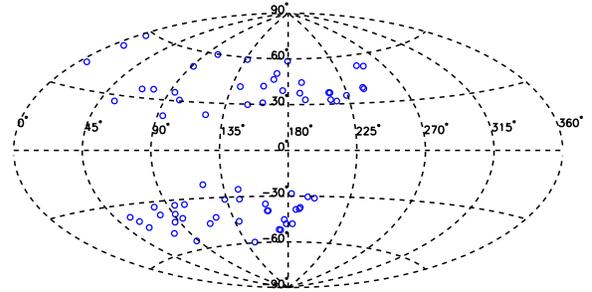}
\caption{The Galactic longitudes and latitudes of 67 BATC observed
fields used in our study. 34 BATC fields belong to south Galactic
latitude field and 33 BATC fields belong to north Galactic latitude
field.} \label{fig1}
\end{center}
\end{figure}

\begin{figure}
\begin{center}
\includegraphics[width=11cm]{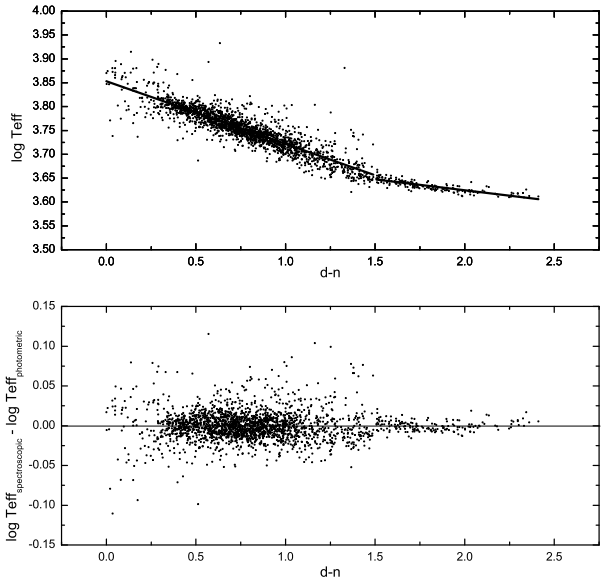}
\caption{The spectroscopical effective temperature function as $d-n$
for 2200 stars is shown in the upper panel, the solid line represent
the effective temperature estimates derived from equation (1). The
bottom panel shows the residuals between the spectroscopic
temperature and photometric temperature estimates.} \label{fig2}
\end{center}
\end{figure}

\subsection{Stellar metallicity estimation}

The most accurate measurements of stellar metallicity are based on
spectroscopic observation. Despite the recent progress in the
availability of stellar spectra (e.g. SDSS--III and RAVE), the
stellar number detected in photometric surveys is much more than
those with spectroscopic observation. So photometric methods have
also often been used to derive the stellar metallicity. For example,
\citet{kar03b} evaluated the metal abundance by ultraviolet-excess
photometric parameter using $UBVI$ data. \citet{nor04} performed a
new fit of the $uvby$ indices to stellar metallicity. \citet{ive08}
obtained the mean metallicity of stars as a function of $u-g$ and
$g-r$ colors of SDSS data. The photometric metallicity methods are
based on the theory that the density of metallicity absorption lines
is highest in the blue spectral region. For the BATC multi-color
photometric system, $a$, $b$, and $c$ filter bands are most
sensitive to stellar metallicity, however, it is unfortunately that
their quantum efficiencies are low. So, the $d-h$ and $d-n$ colors
are used to calibrate stellar metallicity in this paper. As can be
seen from panel (a) of Figure 5, the spectroscopical metallicity of
the 2200 stars as a function of the $d-h$ and $d-n$ shows complex
behavior. The third-order terms are used to derive precise stellar
metallicity estimate. The stellar metallicity estimate as a function
of the $d-h$ and $d-n$ can be found in panel (b) of Figure 5.
\begin{equation}
\begin{split}
%&|x^{2\pi/\omega}(0,t)-x^*-\rho(0)|^2\\
&[Fe/H]=A + B(d-h) +C(d-n) +D(d-h)(d-n) \\
&+E(d-h)^2 + F(d-n)^2 +G(d-h)^2(d-n) +\\
&H(d-h)(d-n)^2 +I(d-h)^3 +J(d-n)^3\\
\end{split}
\end{equation}

There are two sets of coefficients ($A-J$) corresponding to regions
$(d-h) < $ 0.6 ($-$1.62, 0.73, $-$0.35, 3.96, $-$3.60, $-$0.33,
5.89, $-$4.67, 1.21, 0.45) and $(d-h) > $ 0.6 (1.34, $-$10.30, 1.4,
8.37, 7.68, $-$4.47, 13.64, $-$10.87, $-$10.56, 3.20). Metallicities
derived from equation (2) for 2200 stars are compared with the
spectroscopic metallicities. The residuals between the spectroscopic
metallicity and photometric estimate are shown in Figure 6. Stars
within color range $(d-h) < $ 0.6 have residuals of 0.5 dex, while
stars with $(d-h) > $ 0.6 have residuals of 0.3 dex. The metallicity
distribution for 160 000 MS stars in our study is shown in Figure 7.
The peak of stellar metallicity distribution is at $\rm{[Fe/H]} =
-0.7$. The stellar metallicity estimates cover a range of [Fe/H]
from $-1.8$ to $-0.2$.

\subsection{Photometric parallax}

The stellar distances are obtained by photometric parallax relation
which listed in equation (3)
\begin{eqnarray}
m_{v}-M_{v}=5\log r - 5 + A_{v},
\end{eqnarray}
where $m_{v}$ is the visual magnitude, the absolute magnitude
$M_{v}$ can be obtained according to the stellar type which was
derived from stellar effective temperature \citep{du06}. We adopted
the absolute magnitude versus stellar type relation for MS stars
from \citet{lan92}. The reddening extinction $A_{v}$ derived from
\citet{sch98} is small for most BATC survey fields. $r$ is the
stellar distances. The vertical distance of the star from the
Galactic plane can be evaluated by equation (4)
\begin{eqnarray}
z = r sin b,
\end{eqnarray}
where $b$ is the Galactic latitude. A variety of errors affect the
determination of stellar distances. The first source of errors is
from photometric uncertainty and photometric color uncertainty; the
second from the misclassification of stellar type, which should be
small due to the small rms scatter of effective temperature. For
luminosity class V, types F/G, the absolute magnitude uncertainty is
about 0.3 mag. In addition, there may exist an error from the
contamination of binary stars in our sample. We neglect the effect
of binary contamination on distance derivation due to the unknown
but small influence from mass distribution in binary components
\citep{kro93,ojh96}.
\begin{figure}
\begin{center}
\includegraphics[width=9cm]{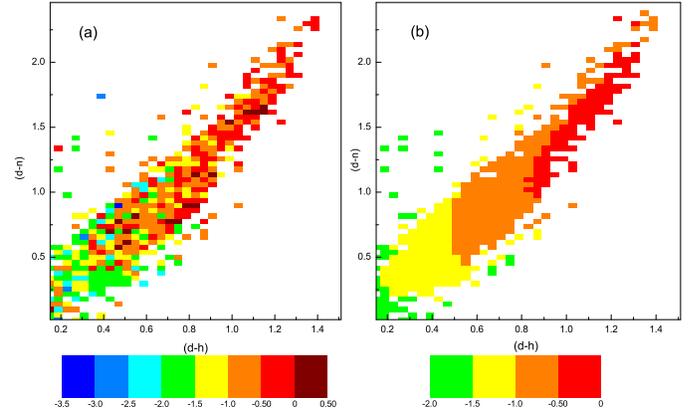}
\caption{The mean [Fe/H] as a function of (d-n) and (d-h) for 2200
stellar spectroscopic metallicity (panel a) and photometric
metallicity estimates (panel b). The color scheme shows the mean
values in all 0.04 mag by 0.04 mag large pixels.} \label{fig6}
\end{center}
\end{figure}

\begin{figure}
\begin{center}
\includegraphics[width=9cm]{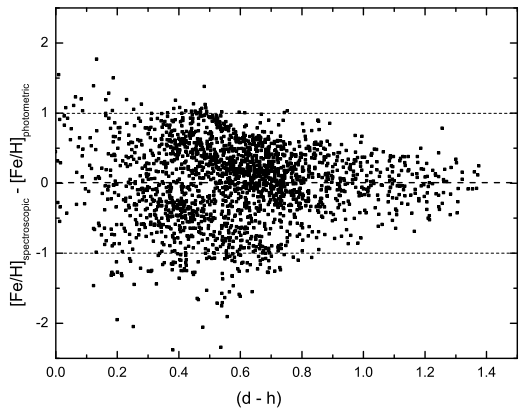}
\caption{The residuals between the spectroscopic metallicity and
photometric metallicity estimates based on equation (2) function as
$d-n$. Stars within color range $d-h < $ 0.6 have rms scatter of 0.5
dex, while stars with $d-h > $ 0.6 have an rms scatter of 0.3 dex.}
\label{fig3}
\end{center}
\end{figure}

\begin{figure}
\begin{center}
\includegraphics[width=9cm]{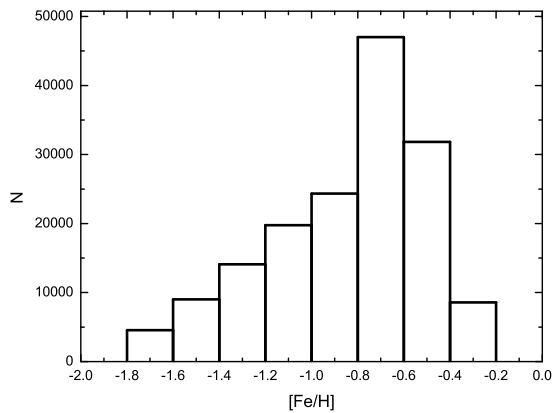}
\caption{The metallicity distribution for 160 000 MS stars.}
\label{fig5}
\end{center}
\end{figure}

\begin{figure}
\begin{center}
\includegraphics[width=9cm]{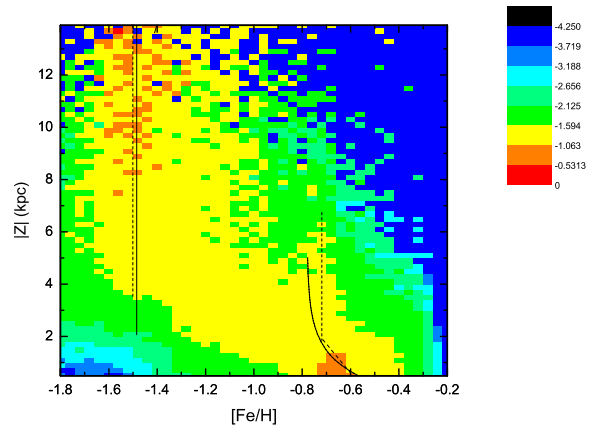}
\caption{The resulting conditional metallicity probability
distribution of 160 000 MS stars for a given Z and $p$([Fe/H]$|$Z).
The color scheme shows the logarithm (base 10) of metallicity
probability.} \label{fig6}
\end{center}
\end{figure}

\section{METALLICITY DISTRIBUTION }

It is well known that the chemical abundance of stellar population
contains much information about the Galaxy formation and evolution.
The study of stellar metallicity distribution in the Galaxy has been
an important subject of photometric and spectroscopic surveys
\citep{per01,kel07,eis11}. In this study, we want to explore
possible stellar metallicity distribution as a function of position.
The metallicity distribution for all MS stars as a function of
vertical distance $|Z|$ is obtained in section 4.1. We first discuss
several features which can be directly found. The detailed behaviors
of metallicity distribution are described in the following sections.

\subsection{The metallicity distribution as a function of vertical distance Z}

According to the sample selected criteria in section 2.1, about 160
000 stars are extracted from 67 fields at median and high latitudes.
Since stellar metallicity distribution at high Galactic latitudes
are not strongly related to the radial distribution \citep{ive08},
they are well suited to study the vertical metallicity distribution
of the Galaxy. The data selected cover a substantial range of $0.5 <
|Z| \leq 14~ \rm{kpc}$. Figure 8 shows the resulting conditional
metallicity probability distribution of 160 000 MS stars for a given
Z and $p$([Fe/H]$|$Z). This distribution is computed as metallicity
histograms in a narrow Z slices and normalized by the total number
of stars in a given Z slice \citep{ive08}. The probability density
derived at an interval of 0.2 kpc in Z distance and 0.04 dex in
metallicity is shown on logarithmic scale.

From Figure 8, it is clear that there is a shift from metal-rich
stars to metal-poor ones with the increasing of vertical distance.
%Two concentrations of stars corresponding to disk ($\rm{[Fe/H]} \sim
%-0.7$) and halo ($\rm{[Fe/H]} \sim -1.5$) stars can be found in
%Figure 6. There is a peak of stellar metallicity distribution at
%$\rm{[Fe/H]} = -0.7$ in the interval $0.5 < |Z| \leq 5~\rm{kpc}$.
From Figure 8, in the interval $|Z| \leq 2~ \rm{kpc}$, most stars
with metallicity richer than $\rm{[Fe/H]} = -1$ belong to the thin
disk stars and the stellar metallicity smoothly decreases with
distance from the Galactic plane. Stars in interval $2 < |Z| \leq 5
~\rm{kpc}$ include both metal-poor and metal-rich stars, there is a
mixture of the halo stars and the thick disk stars. It can be found
that there is a unclear peak of the stellar metallicity distribution
at $\rm{[Fe/H]} = -0.7$ (corresponding to the thick disk). This
result is consistent with that result of \citet{all06}, who found
the thick disc with a maximum at $\rm{[Fe/H]} = -0.7$. From Figure
8, we can see that most stars in interval $|Z| > 5~ \rm{kpc}$ are
metal-poor stars, and the stellar metallicity distribution peaks at
about $-1.5$ which is consistent with the previous work.
\citet{all06} found that the halo stars exhibit a broad range of
metallicity, with a peak at $\rm{[Fe/H]} \simeq -1.4$. \citet{an13}
considered that the metallicity distribution function of the
Galactic halo can be adequately fit using a single Gaussian with a
peak at $\rm{[Fe/H]} = -1.5$. The metallicity of the halo component
appears to be spatially invariant, so there is little or no
metallicity gradient in the halo.

From Figure 8, we can see that the stars in interval $2 <|Z| \leq 5
~\rm{kpc}$ include both metal-poor and metal-rich stars. Stars
exhibit a broad range of metallicity. The metallicity distribution
is mainly in range from $-1.7$ to $-0.5$, indicating that there is a
mixture of the halo and the thick disk components. The metal-rich
stars extend beyond 4--5 kpc from the plane and the stars are
dominated by the halo stars in the interval $|Z| > 5~ \rm{kpc}$
advocating an edge of the Galactic thick disk at about 5 kpc above
and below the plane. Based on our observation, we conclude that the
possible cutoff of the thick disk stars is about 5 kpc above and
below the Galactic plane. \citet{maj92} considered the cutoff of the
Galactic thick disk at about 5.5 kpc above the Galactic plane.
\citet{ive08} analyzed SDSS DR6 photometric data and pointed out
that it seems more likely that the disk is indeed traceable to
beyond 5 kpc from the plane.

\citet{ive08} represented a model for $p$([Fe/H]$|$Z) of the Galaxy
in the range of $0.5 < |Z| \leqslant 10~ \rm{kpc}$. The solid lines
in Figure 8 show the mean metallicity of the disk and the halo for
\citet{ive08}. In their model, the mean halo metallicity
distribution is spatially invariant ($\rm{[Fe/H]} = -1.46$) and the
median metallicity for the disk component in the $0.5 < |Z|
\leqslant 5~ \rm{kpc}$ range can be described as the curve in Figure
8. The dash lines in Figure 8 show the mean metallicity for the halo
and the disk in our study. From Figure 8, we can find that the mean
metallicity of the disk of our study is slightly richer than those
of \citet{ive08} in the range of $2 < |Z| \leqslant 5~ \rm{kpc}$.
However, the mean metallicity of the halo in our study is agree with
the result of \citet{ive08}.

\begin{figure*}
\includegraphics[width=14cm]{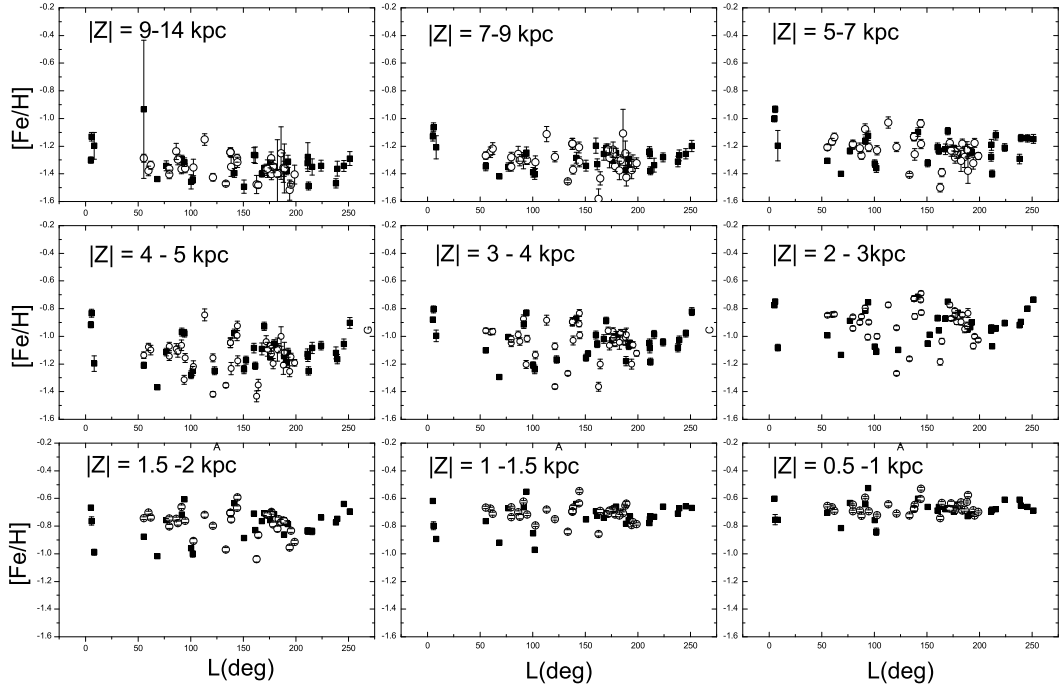}
\caption{The mean metallicity distribution as a function of Galactic
longitude in different distance intervals. Each panel corresponds to
a narrow $|Z|$ range which labeled in each panel.}
\end{figure*}

\begin{figure*}
\includegraphics[width=15cm]{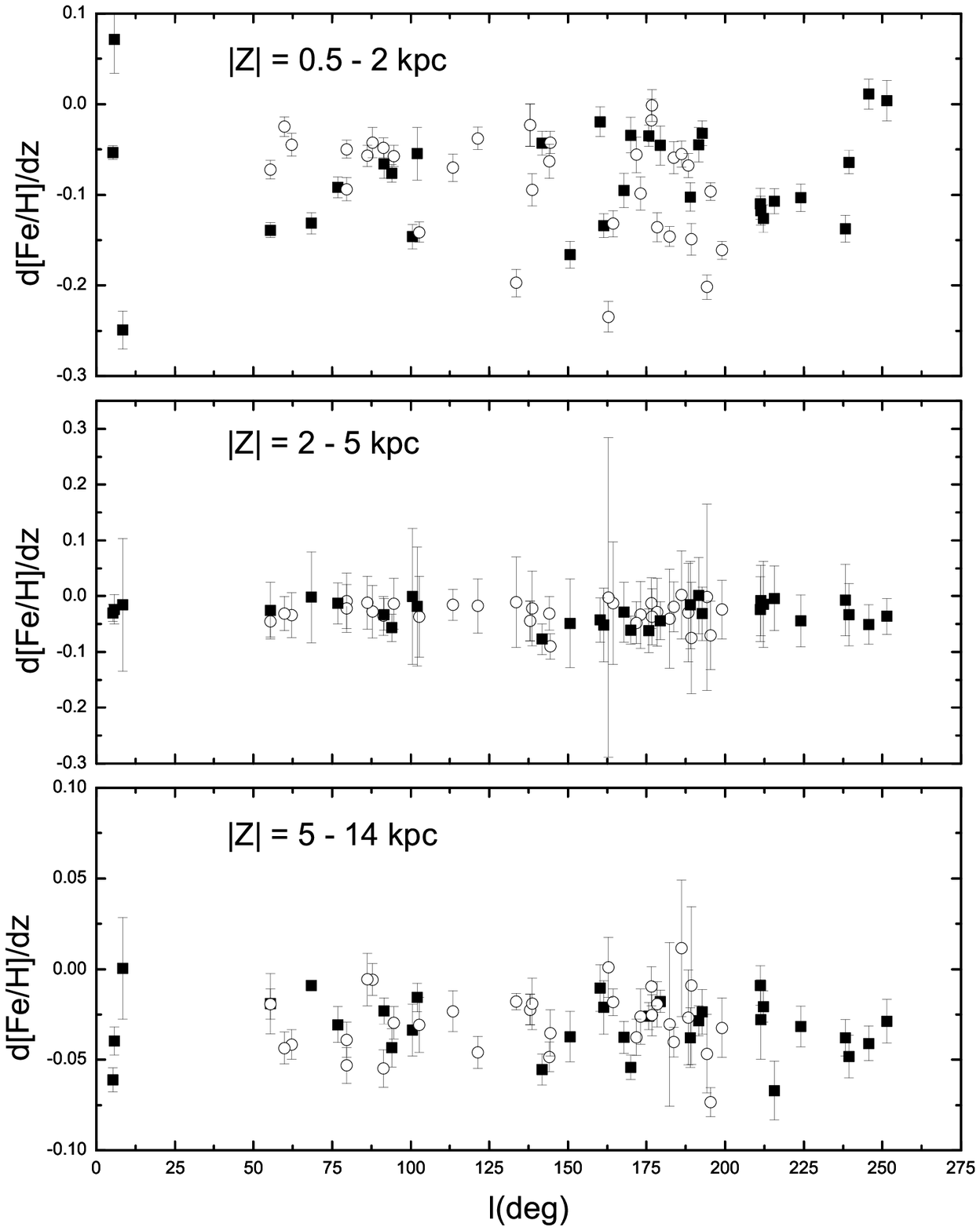}
\caption{The metallicity gradient distribution as a function of
Galactic longitude for the thin disk, the thick disk and the halo
are shown.}
\end{figure*}

\subsection{Metallicity variation with Galactic longitude}

Mean metallicity distribution of each BATC field as a function of
Galactic longitude for different $|Z|$ distance intervals are
presented in Figure 9. The panels from left to right and from top to
bottom correspond to different distances from the Galactic plane,
respectively. The mean metallicity shift from metal-rich to
metal-poor with the increase of distance from the Galactic plane can
be found in Figure 9. The solid points represent the south Galactic
latitude fields, the open square points represent the north Galactic
latitude fields. As shown in Figure 9, at any distance interval, the
variation of the mean metallicity distributions with the Galactic
longitude is almost flat. \citet{ive08} found that the metallicity
distribution of the disk and the halo shifts toward higher
metallicity in the region where affected by the Monoceros stream.
\citet{bil08} found the dependence of the Galactic model parameters
on the Galactic longitude. However, no strong relation between the
mean metallicity and Galactic longitude is found in our work.

The mean metallicity in interval $9 < |Z| \leqslant 14~ \rm{kpc}$ is
around of $-1.34 \pm 0.1 \rm{dex}$. In interval $5 < |Z| \leqslant 9
~\rm{kpc}$, our result indicates that the mean metallicity is $\sim
-1.25$. The mean metallicity decreases from $-0.90$ to $-1.13$ in
intervals $ 2 <|Z|\leqslant 5~ \rm{kpc}$. The mean metallicity
decreases from $-0.65$ to $-0.78$ in intervals $ |Z|\leqslant 2
~\rm{kpc}$. \citet{ive08} found that the median metallicity smoothly
decreases with distance from the Galactic plane from $-$0.6 at 500
pc to $-$0.8 beyond several kpc. The work of \citet{sch12} indicated
that the metallicity distribution function of both G and K dwarf
become more metal-poor with the increasing $|Z|$ from 0.2 to 2.3
kpc.

\subsection{The vertical metallicity gradient}

Detailed information about the vertical metallicity gradient can
provide important clue about the formation scenario of stellar
population. Here, we use the mean metallicity of each BATC field to
derive the vertical metallicity gradient. The metallicity gradients
for all the fields in three different $|Z|$ intervals, $|Z| \leq 2
~\rm{kpc}$ (the thin disk), $2 < |Z| \leqslant  5~ \rm{kpc}$ (the
thick disk) and $5 < |Z| \leqslant 14~ \rm{kpc}$ (the halo), are
given in Figure 10. Compared with the typical error bars, there are
no variations in the gradients with Galactic longitude for the three
different components.

From Figure 10, we can find that the mean gradient of halo is about
$-0.03 \pm 0.02$ dex kpc$^{-1}$ ($5 < |Z| \leqslant 14~\rm{kpc}$),
which is essentially in agreement with the previous conclusions
\citep{du04,ak06,yaz10,pen12}. A metallicity gradient of the halo
($|Z| > 5 ~\rm{kpc}$), $\rm{d[Fe/H]/dz} = -0.05$ dex kpc$^{-1}$, was
found by \citet{pen12}. \citet{yaz10} detected $\rm{d[Fe/H]/dz} =
-0.01$ dex kpc$^{-1}$ for the inner spheroid. \citet{ive08} found
that the halo metallicity distribution is remarkably uniform. No
metallicity gradient found in the halo is consistent with the merger
or accretion origin of the Galactic halo.

At distance $0.5 < |Z| \leqslant 2~ \rm{kpc}$, the mean vertical
abundance gradient d[Fe/H]/dz is about $\sim -0.1 \pm 0.02$ dex
kpc$^{-1}$. This value is consistent with the result of
\citet{ive08}. They detected a vertical metallicity gradient for
disk stars (0.1--0.2 dex kpc $^{-1}$). However, our result is less
than the results of previous works. \citet{pen12} found a
metallicity gradient d[Fe/H]/dz $\sim -0.21 \pm 0.05$ dex kpc$^{-1}$
in the range $|Z| \leq  2 ~\rm{kpc}$. The metallicity gradient was
determined to be d[Fe/H]/dz $\sim -0.37$ dex kpc$^{-1}$ for $ |Z| <
4~ \rm{kpc}$ in the work of \citet{du04}. The range of our
metallicity estimates only cover from $-0.2$ to $-1.8$. The cutoff
of the extreme metal-rich disk stars may result in a little
metallicity gradient in the thin disk.

Recently, the disk of Galaxy has been extensively studied
\citep{ben03,bil12}, especially the thick disk. As shown in Figure
8, a significant mixture of the Galactic components can be found in
the distance interval $2 < |Z| \leqslant 5 ~\rm{kpc}$. For a long
time, the metallicity distribution function of the thick disk was
thought to be cut off below $\rm{[Fe/H]}=-1$. Recently, several
surveys \citep{mor93,bee99,kat99,bee02} argued that the thick disk
includes stars as metal-deficient as [Fe/H] $\sim-1.6$. However, the
metal-weak thick disk appears to be minor constituent of the entire
thick disk. Most metal-poor stars in the thick disk belong to the
halo stars. Here, stars with [Fe/H] $> -1$ in the region $2
<|Z|\leqslant  5~ \rm{kpc}$ are selected to determine the
metallicity gradient for the thick disk. There is a metallicity
gradient $\sim -0.03 \pm 0.02$ dex kpc$^{-1}$ in the thick disk
which is consistent with the result of \citet{all06}. They found
that no vertical metallicity gradient is apparent in this
well-sampled region of the thick disk. The absence of metallicity
gradient in the thick disk favors the conclusion that the thick disk
was formed through the merger or accretion of the nearby dwarf
galaxies.

\section{SUMMARY AND CONCLUSION}

Based on the stellar atmospheric parameters from SDSS spectra for
$\sim2200$ MS stars which were also observed by BATC photometric
system, we develop the polynomial photometric calibration method to
evaluate the stellar effective temperature and metallicity for BATC
photometric data. This calibration method has been applied to about
160 000 MS stars from 67 BATC observed fields. The color index and
magnitude selected criteria are used to choose the MS stars. The
star sample in this paper, 160 000 MS stars, are much more than that
in our previous work \citep{pen12} which used 40 000 MS stars to
derive the metallicity distribution of the Galaxy. The photometric
effective temperature method is capable of deriving effective
temperature with accuracy of 1.3 \% and the photometric metallicity
estimate with error about 0.5 dex.

According to the derived metallicities of all sample stars, we find
that most stars in interval $|Z| > 5~ \rm{kpc}$ are metal-poor stars
which correspond to the halo component. There is a peak of the
stellar metallicity distribution at about $-1.5$ in this interval.
The halo metallicity distribution seems invariant with the distance
from the plane while the mean thin disk (0.5 $< |Z| \leq  2
~\rm{kpc}$) metallicity smoothly decreases from $-0.65$ to $-0.78$
with distance from the Galactic plane. In interval $2 < |Z|
\leqslant 5~ \rm{kpc}$, there is a mixture of the halo stars and the
thick disk stars. The peak of the metallicity distribution for the
thick disk is about $-0.7$. We believe that the possible cutoff of
the thick disk stars is about 5 kpc above and below the Galactic
plane. At any distance interval, the variation of the mean
metallicity distributions with Galactic longitude is almost flat.

There is a mean vertical metallicity gradient d[Fe/H]/dz = $-0.1$
dex kpc$^{-1}$ for the thin disk stars ($0.5 < |Z| \leqslant2
~\rm{kpc}$). Stars with [Fe/H] $> -1$ and $2 <|Z|\leqslant 5~
\rm{kpc}$ are selected to determine the metallicity gradient for the
thick disk. There is a vertical metallicity gradient $\sim -0.03 \pm
0.02$ dex kpc$^{-1}$ in the thick disk. Meanwhile, the mean vertical
gradient of the halo is about $-0.03 \pm 0.02$ dex kpc$^{-1}$ ($5 <
|Z| \leqslant 14 ~\rm{kpc}$). The absence of the vertical gradient
in the thick disk and the halo favor the conclusion that the thick
disk and the halo may be formed through merger or acceration. The
variations of the vertical gradient for different Galaxy components
with Galactic longitude are flat.

\section*{ACKNOWLEDGMENTS}
We especially thank the anonymous referee for numerous helpful
comments and suggestions that have significantly improved this
manuscript. This work was supported by the Chinese National Natural
Science Foundation grant Nos. 11073032. This work was also supported
by the joint fund of Astronomy of the National Nature Science
Foundation of China and the Chinese Academy of Science, under Grants
U1231113.

\end{document}